\documentclass[aps,prb,twocolumn,showpacs,floats,floatfix]{revtex4-1}

\usepackage{graphicx}
\usepackage{amsmath}
\usepackage{amssymb}
\usepackage{bm}

\usepackage{color}

\begin{document}

\title{Degeneracy doubling and sublattice polarization in strain-induced pseudo-Landau levels}

\author{Charles Poli}
\affiliation{Department of Physics, Lancaster University, Lancaster, LA1 4YB, United Kingdom}
\author{Jake Arkinstall}
\affiliation{Department of Physics, Lancaster University, Lancaster, LA1 4YB, United Kingdom}
\author{Henning  Schomerus}
\affiliation{Department of Physics, Lancaster University, Lancaster, LA1 4YB, United Kingdom}
\date{\today}

\begin{abstract}
The degeneracy and spatial support of pseudo-Landau levels (pLLs) in strained honeycomb lattices systematically depends on the geometry -- for instance, in hexagonal and rectangular flakes the 0th pLL displays a twofold increased degeneracy, while the characteristic sublattice polarization of the 0th pLL is only fully realized in a zigzag-terminated triangle.
These features are dictated by algebraic constraints in the atomistic theory, and signify a departure from the standard picture in which all qualitative differences between pLLs and Landau levels induced by a magnetic field trace back to the valley-antisymmetry of the pseudomagnetic field.
\end{abstract}
\maketitle

\section{Introduction}

In graphene \cite{Cas09,Gui09, Per09a, Lev10, Low10, Lu12, Yan12, Klimov22062012} and chemically functionalised or patterned electronic and photonic analogues with an underlying honeycomb lattice,  \cite{Gom12, Sza12, Bel13, Rec13, Sch13} inhomogeneous strain
influences motion in a manner similar to an effective
magnetic field. \cite{Cas09,Gui09,Ior85,Kan97,Suz02} In the conventional continuum approximation, this equivalence arises since the strain adds a position-dependent term to the momentum operator in the Dirac Hamiltonian, acting very much like a vector potential; \cite{Voz10} the only difference to a real magnetic field is the fact that the pseudomagnetic field is opposite near the two inequivalent K points (valleys) in the Brillouin zone of the unstrained system. In a strain configuration which corresponds to a constant pseudomagnetic field one therefore expects a standard sequence of Landau levels, then termed pseudo-Landau levels (pLLs), but edge states are non-chiral while the 0th pLL should display an identical sublattice polarization in both valleys (see also Refs.~\onlinecite{Gui10, Pra10, Gop12, Nee13, Cos13, Qi13}).

A consequence of the stated equivalence is the expectation that the degeneracy of the pLLs obeys the standard filling-factor rules of the quantum Hall effect \cite{novoselovtwo2005,Zha06,Goe11}
and thus
universally depends on the area of the system, with only little influence of the geometry. \cite{Bre06, Rom11, Zar11, Guc13} Here we point out that contrary to this expectation, the degeneracy---as well as other characteristic properties of pLLs, such as sublattice polarization and the support of the pLLs in the bulk and at the edges---depends systematically on the geometry.
In particular, in hexagons and rectangles, the 0th pLL displays a doubled degeneracy, with all states having an equal weight on both sublattices---the sublattice polarization in the bulk is balanced exactly by an opposite polarization at the edges of the system.
The characteristic sublattice polarization in the 0th pLL is only fully realized in a particular geometry, the zigzag-terminated triangle.
Notably, truncating the triangle to the hexagon (which reduces the area by 2/3) \emph{increases} the degeneracy the 0th pLL (to three times the value expected from the area argument), while the degeneracy of the other pLLs remains unaffected (thus also not obeying the area argument).
This behaviour
is dictated by the physical conditions for the formation of pLLs (which require large strain) and strict algebraic constraints in the atomistic description (limiting the number of sublattice-polarised zero-energy states in finite systems), which conspire to yield a systematic hybridization of bulk and edge modes not captured in the low-energy continuum theory.

Section \ref{sec:cond} describes the optimal conditions for a fully  established sequence of strain-induced pseudo-Landau levels. Section \ref{sec:uncon} describes  unconventional features that set these states apart from Landau levels induced by a magnetic field.  Section \ref{sec:relation} exploits the strict algebraic constraints to establish the relation between the degeneracy of the 0th pLL and its sublattice polarization. Our results are summarised in Sec.~\ref{sec:conclusions}.

  \begin{figure*}[t]
\includegraphics[width=.8\textwidth]{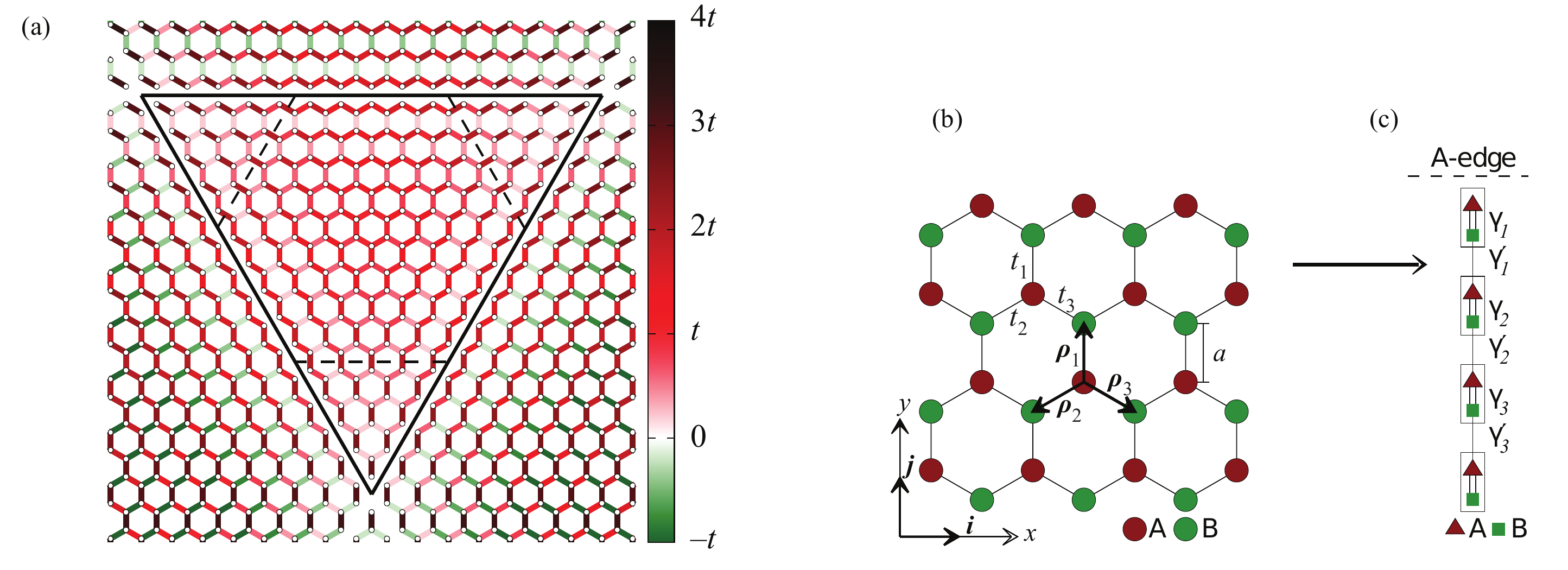}
 \caption{(Color online)
(a) Strain-induced coupling pattern corresponding to a constant pseudomagnetic field.
The region of positive couplings maps out a zigzag-terminated triangle, which defines the optimal geometry to realise a sublattice-polarised 0th pLL, see Fig.~\ref{zerothLL}(a). The dashed lines indicate the truncation of the triangle to a hexagon, by which the degeneracy in the 0th pLL doubles while the sublattice polarization disappears, see Fig.~\ref{zerothLL}(b).
(b) The underlying honeycomb lattice is composed of the sublattices A and B, connected by bond vectors $\boldsymbol{\rho}_l$, $l=1, 2, 3$. (c) Mapping onto an effective dimer chain, used at the end of this paper.
}\label{lattice}
\end{figure*}

\section{Conditions for formation of pseudo-Landau levels}
\label{sec:cond}

We start by discussing the conditions under which pLLs are clearly formed in a strained electronic or photonic honeycomb lattices.  As the principal features  investigated here are robustly protected by the energy gaps, we neglect higher-order corrections and the deformation potential that would apply to the case of specific graphene. \cite{Rib09,Kit12,deJ12}
Within the low-energy theory,
the effective Dirac Hamiltonian in the presence of strain then reads
\begin{equation}
H=v\eta \sigma_x( p_x-A_x)+v\sigma_y( p_y-A_y),
\end{equation}
where $v=3ta/2\hbar$ is the Fermi velocity  in the unstrained lattice with bond length $a$ and coupling constant $t$. Furthermore, $\sigma_{x,y}$ are the Pauli matrices in the A/B sublattice space, $p_{x,y}=-i\partial_{x,y}/\hbar$, and $\eta=\pm 1$ is the valley index. The effective vector potential
\begin{eqnarray}
A_x&=&\eta(2 t_1-t_2-t_3)/(3 at),
\\
A_y&=&\eta(t_2-t_3)/(\sqrt{3} at)
\end{eqnarray}
is valley-antisymmetric and depends on the local values of the couplings $t_l$
along each of the three bond orientations $\boldsymbol{\rho}_l$, $l=1,2,3$  in the honeycomb lattice, see Fig.~\ref{lattice}(b).
The largest value of the pseudomagnetic field is obtained for a triaxial strain configuration, \cite{Gui09} in which the underlying tight-binding couplings
\begin{equation}
t_l=t[1-(\beta/2)\boldsymbol{\rho}_l\cdot {\bf r}_l]
\end{equation}
depend linearly on the bond centre position ${\bf r}_l$ along each of the three bond orientations, see Fig.~\ref{lattice}(a). Here $\beta$ corresponds to the strength of the pseudomagnetic field (in units of $\hbar c/e$), and the pseudomagnetic length is given by $\ell=\sqrt{1/|\beta|}$. The pLLs have energies
\begin{equation}
E_n={\rm sgn}(n)\hbar v\sqrt{2|\beta n|}\quad (n\mbox{~integer}).
\end{equation}
The 0th pLL is localized on the A sublattice if $\beta>0$ and on the B sublattice if $\beta<0$, while all other pLLs have equal weight on both sublattices.

\begin{figure*}[t]
\includegraphics[width=\textwidth]{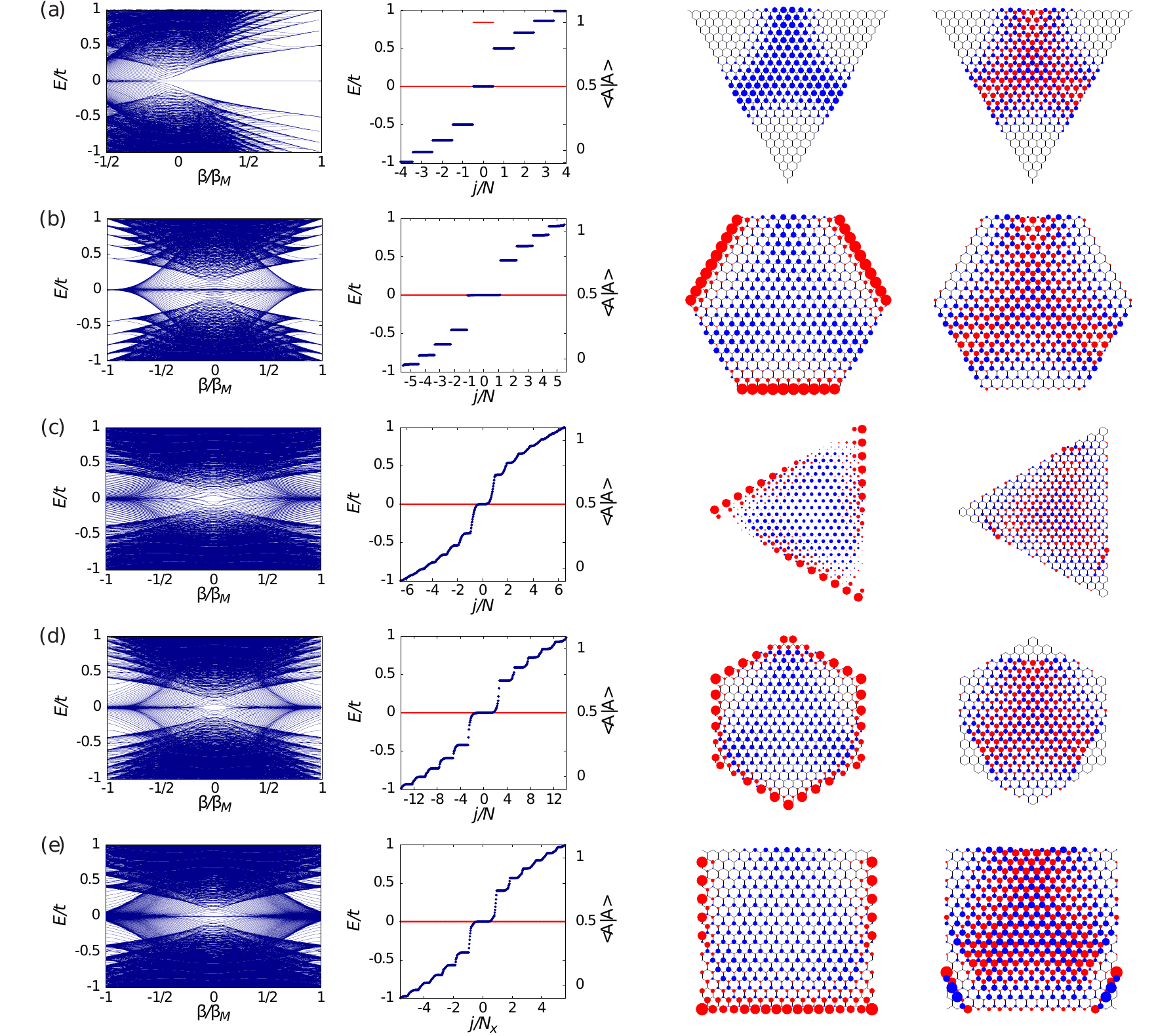}
 \caption{(Color online) Energies and wavefunction support of states in strained honeycomb flakes. (a) Zigzag-terminated triangle, (b) zigzag-terminated hexagon, (c) armchair-terminated triangle, (d) armchair-terminated hexagon, (e) approximate square. The
first column shows the dependence of the energy levels of the pseudomagnetic field $\beta$, with $\beta_M$ the maximal value at which couplings stay positive throughout a given system.
In the second column, the dark blue dots show the energy level staircase at $\beta=\beta_M$ (left axis) while the brighter red dots show the weight of eigenstates on the A sublattice (right axis).
The third and fourth column show the total probability density of the states associated to the 0th and 1st pLL (dark blue on A sites and bright red on B sites).
The system sizes (number of edge atoms) for first and second (third and fourth) columns are (a) $N=74$ ($N=24$), (b) $N=30$ ($N=10$), (c) $N=84$ ($N=26$), (d) $N=36$ ($N=12$), (e) $N_x=49$, $N_y=56$ ($N_x=17$, $N_y=18$).
 }\label{zerothLL}
\end{figure*}

When we implement the triaxial strain in finite systems
we find that the pLLs are only clearly formed at strains where couplings $t_l\ll t$ become very small close to some edges of the system. This is shown in Fig.~\ref{zerothLL} for a number of representative geometries (triangles and hexagons with zigzag and armchair edges, and an approximate square), where the left panels display the strain-dependence of the low-lying energy levels (analogous to Hofstadter's butterfly for systems exposed to real magnetic fields). In these panels, $\beta_M$ signifies the strain at which couplings at the edge of the system drop to zero. In all cases, the pLL sequence only clearly forms for values of $\beta$ approaching $\beta_M$.
As the couplings should remain positive,
these considerations also define a natural limiting size of a strained flake with a given uniform pseudomagnetic field---in a bulk honeycomb lattice, the lines where couplings change sign in the triaxial strain profile cut out a zigzag-terminated triangle, shown in Fig.~\ref{lattice}(a), which therefore is the optimal geometry to realize strain-induced pLLs.

We also implemented triaxial strains with different distance dependence of the couplings, e.g., an exponential dependence which guarantees that all couplings remain positive, but this result in a much increased dispersion making it difficult to identify the pLLs.
Therefore, the linear strain profile remains optimal even when one goes beyond the continuum approximation.

\section{Unconventional features of pseudo-Landau levels}
\label{sec:uncon}

The constraints on size and strain identified in the previous section result in a large variation of coupling strengths across the system. Therefore, it can be anticipated that
pLLs inherit properties which go beyond the simple equivalence to real magnetic fields predicted by the standard low-energy theory.  We identify three such features---the degeneracy of the pLLs, the sublattice polarization, and the support of the wave functions in the bulk and at different types of edges. The remaining panels in
Fig.~\ref{zerothLL} give an overview of these properties for the mentioned representative systems. The second column shows the energy level staircase (left axis) and weight of the states on the A sublattice (right axis), while the two rightmost panels show the total probability densities of the states in the 0th and 1st pLL. These results are obtained for the maximally allowed strain value $\beta_M$ in each given geometry.

Figure \ref{zerothLL} (a) is for the triangle with zigzag edges, the optimal geometry identified above. Here, the level staircase is fully developed, showing only minimal dispersion in all pLLs (the remaining amount of dispersion diminishes when the system size is increased). The degeneracy of the levels, obtained from the width of the steps in the level staircase, is given by $N-|n|$, where $N$ is the number of A atoms along each edge of the triangle. In this geometry the 0th pLL displays the expected sublattice polarization, with all the weight concentrated on the A sublattice, while the states in the 1st pLL have equal weight on both sublattices. The probability density in these pLLs avoids the areas close to the corners, and instead maps out a hexagonally shaped area.

While these properties of the zigzag-terminated triangle broadly conform with the expectations from the standard low-energy theory, the results for the remaining geometries in  Figure \ref{zerothLL}  demonstrate that this an exception. In particular, we generally find a systematic enhancement of the degeneracy of the 0th pLL, which goes along with a loss of the sublattice polarization originating from a hybridization with edge states localized on the B sublattice.
Part (b) shows the results for a hexagonal flake with zigzag edges, where three edges are terminated by $N$ A atoms while the others are terminated by $N$ B atoms. The degeneracy of the 0th pLL is now given by $6N$, while the other pLLs are composed of $3N-|n|$ levels. Furthermore, we now observe that
\emph{all} states have an exactly equal weight on both sublattices, with the weight on the B sublattice concentrated at the three edges terminated by B atoms.
In the presence of armchair edges (parts c and d for the triangle and the hexagon), the spectral gaps between the pLLs become filled with additional states, which persist even at the maximal value of the strain. The strain-dependent energy levels in the butterfly form clear caustics, which allow to attribute each state to a pLL.
In leading order of the  number $N$ of atoms along each of the armchair edges of the system (counting both A and B sites), the zeroth pLL in the armchair triangle is $2N$-fold degenerate, while the other pLLs are $N$-fold degenerate; for the armchair hexagon, the  0th pLL is $6N$-fold degenerate while the other pLLs are $3N$-fold degenerate. In both cases, the 0th pLL does not display any sublattice polarization, with the weight on the B sublattice concentrated  on the edges and distorted towards the regions where the couplings along the edge bonds are weak. The same general observations hold for rectangular flakes  (which both have armchair and zigzag edges), as shown for the approximate square in part (e). Here, in leading order the degeneracy of the 0th pLL is given by $2N_x+1.5 N_y$,
while that of the other pLLs is again halved ($N_x$ is the number of A atoms along the zigzag top edge or B atoms along the zigzag bottom edge, while $N_y$ is the number of A and B atoms along each armchair edge).
The counterweight on the B sublattice is distorted towards the bottom edge, as well as towards the top of the armchair edges.

\section{Relation between degeneracy and sublattice polarization}
\label{sec:relation}

The numerical results suggest a strong link between the degeneracy of the pLLs and their sublattice polarization.
We now establish this connection for general geometries, leading to a picture that we can verify for a sequence of appropriately designed shapes. This picture relies on a combination of strict algebraic constraints (relating to the excess of A atoms over B atoms) and a closer inspection of the physics at the edges of the system, which are the source of low-energy states that  hybridize with the strain-induced bulk states.

The first step of the argument concerns the number of sublattice-polarized zero modes in bipartite systems with simply connected geometry. \cite{PhysRevB.34.5208,Lieb89,PhysRevB.49.3190,PhysRevB.66.014204} In such systems, the Hamiltonian takes a block form where the finite entries make up a rectangular matrix $H_{AB}$ of dimensions $n_A\times n_B$, given by the total number of A and B atoms, respectively. For this structure of the Hamiltonian, there are then $|n_A-n_B|$ states at exact zero energy (a statement known as Lieb's theorem), and these are all fully sublattice polarized (on the A sublattice if $n_A>n_B$ and on the $B$ sublattice if $n_B>n_A$). Furthermore, it follows from a simple algebraic argument that all other states have equal weight on both sublattices ($\langle A|H_{AB}| B\rangle=E\langle A|A\rangle=E\langle B| B\rangle$, where $|A,B\rangle$ are the state vectors in the A and B subspaces). For the optimally strained zigzag triangle, $|n_A-n_B|=N$, so that all states in the zeroth pLL are of this origin, while for all the other geometries discussed so far $n_A=n_B$, meaning that they do not admit any sublattice-polarized zero states.

These algebraic constraints leave the question of the origin and (somewhat paradoxically) \emph{increased} degeneracy of the non-polarised states in the zeroth pLL. According to the observed wave function support, these states should arise from a hybridization of the bulk pLL states predicted from the low energy theory, and low-energy edge states localized on the B sublattice.
Here, we have to distinguish armchair edges and two types of zigzag edges, where the latter are either terminated by A or by B atoms.
In terms of the strain-induced couplings, zigzag edges terminated by A atoms are weakly coupled to the rest of the system; an example are the edges of the optimally strained triangle.
In contrast, zigzag edges terminated by B atoms are strongly coupled to the rest of the system; an example are the three edges created by truncating the zigzag triangle to a hexagon.
Furthermore, along an armchair edge, the coupling strengths directed in parallel to the edge increase linearly as one moves along one direction.

Based on these coupling patterns, the wavefunction weights can now be inferred by separating the couplings along and perpendicular to an edge, leading to an effective one-dimensional model as indicated in Fig.~\ref{lattice}(c).
Along a zigzag edge terminated by A sites and aligned along the $x$ axis, the couplings $t_2,t_3=O(t)$ vary quasi-continuously unless one approaches the corners of the optimally strained zigzag triangle. In perpendicular direction, the system can be viewed as strands of A and B sites, represented by amplitudes $\phi^{(A)}_n$ and $\phi^{(B)}_n$, with alternating couplings $\gamma_n$ arising from $t_{2,3}$ and $\gamma'_n$ arising from $t_1$. Locally, the behaviour into this direction can therefore be approximated by a modulated one-dimensional dimer chain, \cite{Su79}
\begin{eqnarray}
E\phi^{(A)}_n&=&\gamma_n \phi^{(B)}_n+\gamma_{n-1}' \phi^{(B)}_{n-1}, \\
E\phi^{(B)}_n&=&\gamma_n \phi^{(A)}_n+\gamma_{n}' \phi^{(A)}_{n+1}.
\end{eqnarray}
This description is exact for a uniaxial strain configuration in which $t_2=t_3=t$, for which the problem is separable (in the continuum approximation, this simply corresponds to a different gauge). In this case $\gamma_n=2t\cos(\sqrt{3}k_xa)$, where $k_x$ is the wave number in $x$ direction, while $\gamma_n'=t_1$.

\begin{figure*}[t]
\includegraphics[width=\textwidth]{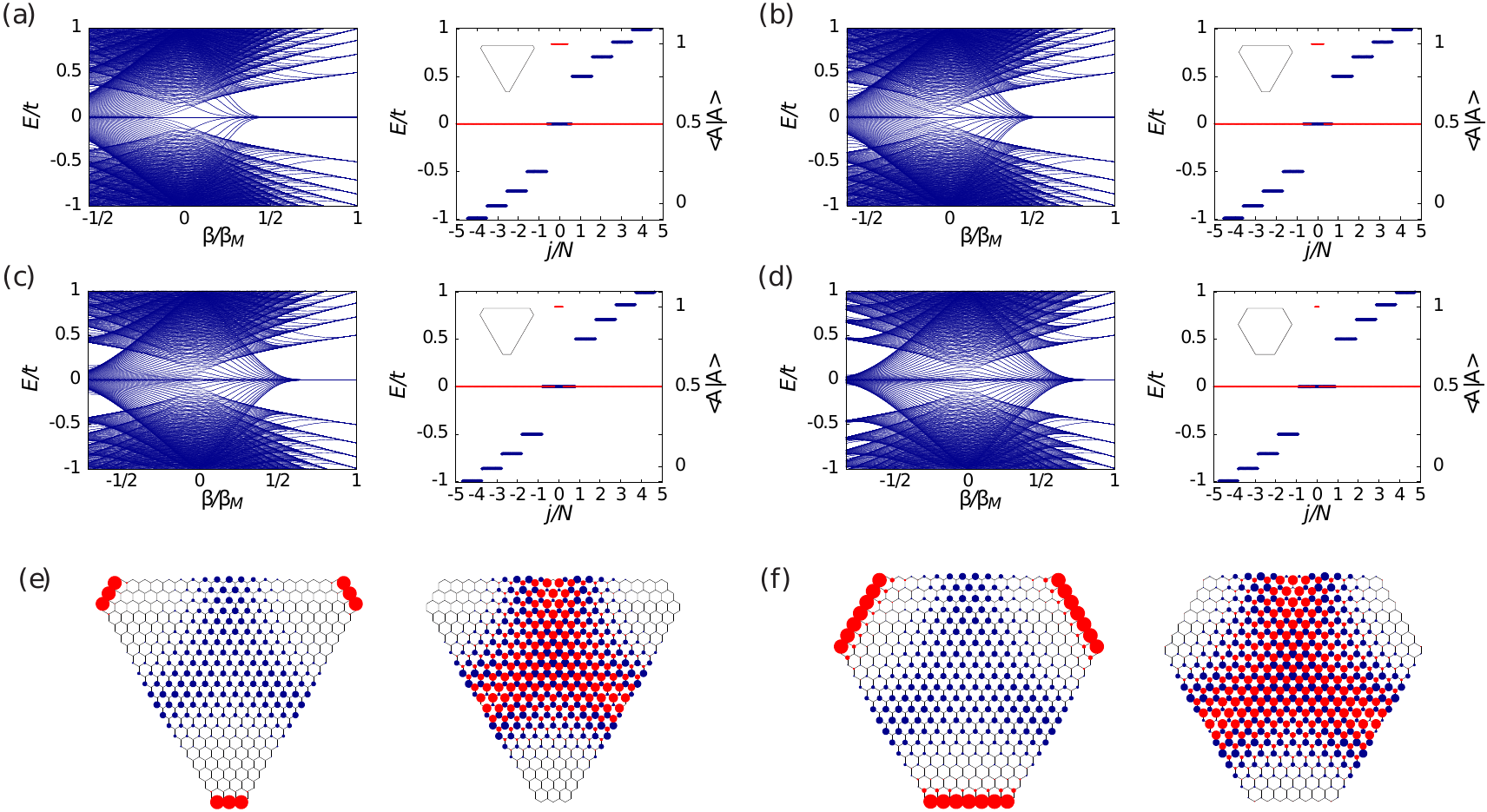}
 \caption{(Color online) Energies (a-d) and wavefunction support (e,f)  of states in strained honeycomb flakes, for several examples of truncated zigzag-terminated triangles [interpolating between the untruncated zigzag-terminated triangle in Fig.~\ref{zerothLL}(a) and the zigzag-terminated hexagon in Fig.~\ref{zerothLL}(b)].
 In (a-d), the first panel shows the strain dependence of energies while the second panel shows the
 energy level staircase at $\beta=\beta_M$ (left axis) and the weight of eigenstates on the A sublattice (right axis). The outline of the flake is indicated in the inset. The full triangle has size $N=74$ and then is truncated at the corners by removing $m$ rows of edge atoms (a: $m=5$, b: $m=10$, c: $m=15$, d: $m=20$). In (e) and (f), the first panel shows the total probability density of the states associated to the 0th  pLL  (dark blue on A sites and bright red on B sites), while the second panel shows the analogous result for the 1st  pLL. (e: $N=25$, $m=3$; f: $N=28$, $m=7$).
 }\label{fig:truncated}
\end{figure*}

Starting from the dimer chain we see that as in unstrained graphene, zero-energy states
\begin{equation}
\phi^{(A)}_n=\phi^{(A)}_1\prod_{1\leq m<n} (\gamma_m/\gamma'_m)
\end{equation}
reside on the A sublattice. However, the strain
increases the factors $(\gamma_m/\gamma'_m)$,
meaning that the wave functions are moved into the bulk. This expulsion effect is strongest close to the corners, resulting in the approximately hexagonally shaped support of the pLLs observed in Fig.~\ref{zerothLL}(a).

The effective dimer chain also applies to B-terminated zigzag edges, with the role of the sublattices interchanged. These edges thus provide a source of low-energy states localized on the B sublattice. Due to the different coupling pattern, the associated wave functions now experience an increased localization at the edges, to the extent that states which would be deconfined in the unstrained system now become well confined. This not only explains the pronounced enhancement of these states at such edges, but also means that  for the zigzag-terminated hexagon [Fig.~\ref{zerothLL}(a)] a sufficient number ($3N$) of states  can be generated to hybridize with the $3N$ bulk states predicted by the low-energy theory. The general algebraic considerations mentioned above then enforce a hybridization into $6N$  states that have exactly equal weight on both sublattices.

For armchair edges, the effective dimer chain  applies for the direction along the edges, which induces the systematic modulation of the probability density seen in Fig.~\ref{zerothLL}(c)-(e). However, these edges does not provide a source of low-energy  states, forcing higher-energy states into the hybridization, which explains why the gaps between the pLLs are now filled.

We tested these general considerations for the transition from a zigzag triangle to the zigzag hexagon, generalising the situation sketched in the introduction. For this we truncate the corners of a triangle of size $N$, thereby creating three additional edges terminated by $m$ B atoms; $m=0$ corresponds to the original triangle while $m=N/3$ corresponds to the hexagon. Despite reducing the area, we then indeed find an increasing number of $N+3m$ modes in the 0th pLL, of which $N-3m$ show full sublattice polarization while $6m$ have equal weight on both sublattice. The degeneracy of the other pLLs remains fixed at $N-|n|$. This is shown for some examples in Fig.~\ref{fig:truncated}.

\section{Conclusions}
\label{sec:conclusions}

In conclusion, key features of pseudo-Landau levels (pLL) in strained honeycomb systems are governed by geometry-dependent physics which goes beyond the simple equivalence with real magnetic fields implied by the standard low-energy continuum theory. In particular, in typical geometries such as hexagons and rectangles, the 0th pLL displays a doubled degeneracy. This originates from the hybridization of bulk states localized on the A sublattice with edge states localized on the B sublattice and results in a loss of sublattice polarization,  as dictated by algebraic constraints in the atomistic theory.

It is noteworthy that the non-universal features described here already arise under the most ideal circumstances (identified in Sec.~\ref{sec:cond}).
Our results show that the system indeed has to be stretched to the physical limits to achieve a well-defined sequence of pLLs. Applications intending to exploit the sublattice polarization of the 0ths pLL should aim to realize the geometry of an optimally strained zigzag triangle.

Various technologies exist to engineer strain configurations and edges in graphene \cite{Low10, Lu12, Li08, Campos09, hicksa2012} as well as
in patterned electronic and photonic realisations of honeycomb systems. \cite{Gom12, Sza12, Bel13, Rec13}
Depending on the physical platform, additional obstructions occur in the form of bulk or edge disorder, or higher-order couplings that serve as additional sources of asymmetry and dispersion.

\acknowledgments

We gratefully acknowledge discussions with Pablo San-Jose, Diana Cosma and Vladimir Fal'ko, as well as the support of EPSRC via grant EP/J019585/1.


%

\end{document}